\documentclass[journal,a4paper,doublecolumn,10pt]{IEEEtran}
\IEEEoverridecommandlockouts
\usepackage{amssymb}
\usepackage{amsfonts}
\usepackage{epstopdf}
\usepackage{graphicx}
\usepackage{stmaryrd}
\usepackage{amssymb,amsmath}
\usepackage{multirow,url}
\usepackage{color,cite}
\usepackage{verbatim} 

\ifodd 1
\newcommand{\com}[1]{\textbf{\color{red} (COMMENT: #1)}} 
\newcommand{\comg}[1]{\textbf{\color{green} (COMMENT: #1)}}
\newcommand{\response}[1]{\textbf{\color{magenta} (RESPONSE: #1)}} 
\else

\newcommand{\com}[1]{}
\newcommand{\comg}[1]{}
\newcommand{\response}[1]{}
\fi
\ifodd 0
\newcommand{\referred}[1]{\textcolor{red}{RefPaper: #1}} 
\else
\newcommand{\referred}[1]{}
\fi

\ifodd 1
\newcommand{\changeblue}[1]{\textcolor{blue}{Modified: #1}} 
\else
\newcommand{\changeblue}[1]{}
\fi

\begin{document}

\title{Multiuser Rate-Diverse Network-Coded Multiple Access}

\author{Haoyuan Pan, Lu Lu, and Soung Chang Liew \\
Department of Information Engineering, The Chinese University of Hong Kong, Hong Kong.\\
Emails:\{ph014, lulu, soung\}@ie.cuhk.edu.hk
}

\maketitle

\begin{abstract}
This paper presents the first Network-Coded Multiple Access (NCMA) system with multiple users adopting different signal modulations, referred to as \emph{rate-diverse} NCMA. A distinguishing feature of NCMA is the joint use of physical-layer network coding (PNC) and multiuser decoding (MUD) to boost throughput of multipacket reception systems. In previous NCMA systems, users adopt the same modulation regardless of their individual channel conditions. This leads to suboptimal throughput for many practical scenarios, especially when different users have widely varying channel conditions. A rate-diverse NCMA system allows different users to use modulations that are commensurate with their channel conditions. A key challenge is the design of the PNC mapping and decoding mechanisms in NCMA when different users adopt different modulations. While there have been past work on non-channel-coded rate-diverse PNC, this paper is the first attempt to design channel-coded rate-diverse PNC to ensure the reliability of the overall NCMA system. Specifically, we put forth a symbol-splitting channel coding and modulation design so that PNC/NCMA can work over different modulations. We implemented our rate-diverse NCMA system on software-defined radios. Experimental results show that the throughput of rate-diverse NCMA can outperform the state-of-the-art rate-homogeneous NCMA by 80\%. Overall, the introduction of rate diversity significantly boosts the NCMA system throughput in practical scenarios.

\end{abstract}


%
\IEEEpeerreviewmaketitle



\section{Introduction}
\begin{figure}
\centering
\includegraphics[width=0.35\textwidth]{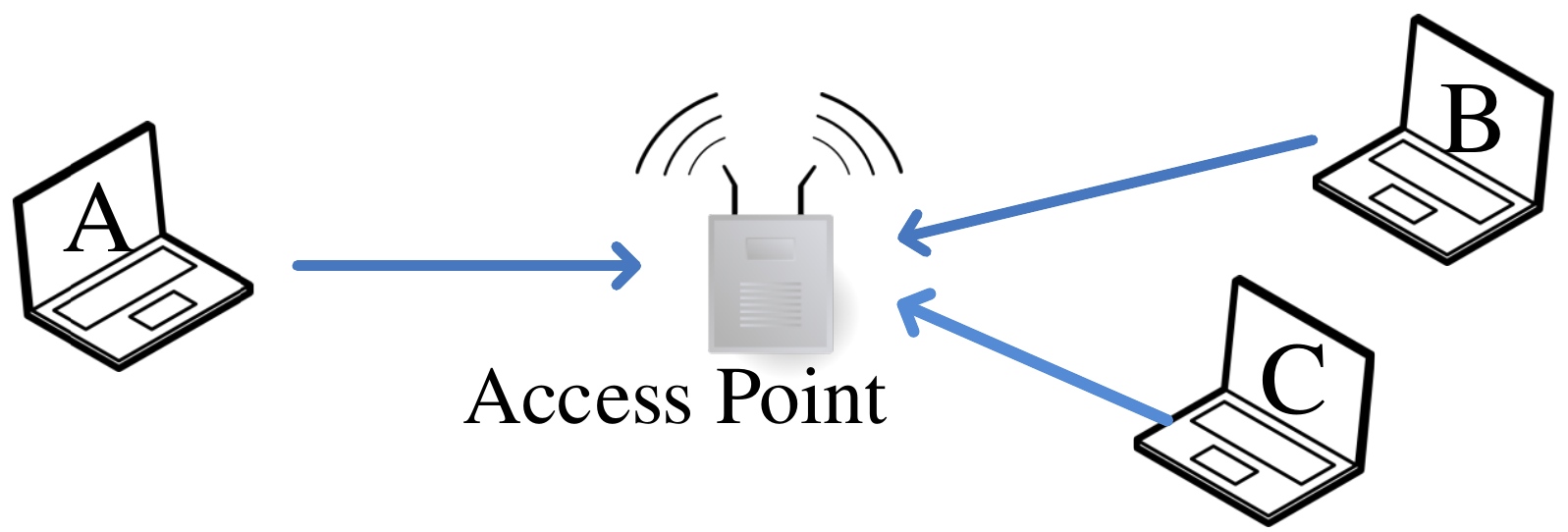}
\caption{NCMA system model with three end users.}\label{fig:system_model}
\vspace{-0.2in}
\end{figure}

This paper studies Network-Coded Multiple Access (NCMA) systems with multiple users adopting different signal modulations, referred to as \emph{rate-diverse} NCMA. NCMA, first proposed in \referred{NCMA1}\cite{NCMA1}, is a new Non-Orthogonal Multiple Access (NOMA) \referred{NOMAfor5G,NOMAVTC13}\cite{NOMAfor5G,NOMAVTC13} architecture with multipacket reception capability. Fig. \ref{fig:system_model} shows a typical wireless local area network (WLAN) in which three end users send messages to a common access point (AP). To boost throughput, the three users are allowed to send their packets simultaneously.

Conventionally, multipacket reception is realized by multiuser decoding (MUD) techniques. The key idea of NCMA is to combine physical-layer network coding (PNC) and MUD to enable multipacket reception. PNC, first introduced in \referred{PNC06}\cite{PNC06}, turns mutual interference between signals from simultaneous transmitters to useful network-coded information, thereby improving the throughput of wireless relay networks. Most prior PNC works focused on relay networks, while NCMA was the first multiple access scheme that explored the use of PNC decoding for non-relay networks, e.g., uplink of WLAN \referred{NCMA1}\cite{NCMA1}. MUD, on the other hand, has been widely studied in the past few decades, from orthogonal signaling (e.g., TDMA, CDMA and OFDMA) to non-orthogonal signaling (e.g., NOMA) \referred{NOMAfor5G,Verdubook}\cite{NOMAfor5G,Verdubook}.

In NCMA, each end user (e.g., user A, B or C in Fig. \ref{fig:system_model}) partitions and encodes one large source message to multiple small packets at the MAC layer (see Fig. \ref{fig:general_architec} \referred{NCMA1}\cite{NCMA1}. At the PHY layer, additional channel coding is performed on each small packet before it is transmitted to the AP. At the PHY layer of the AP, PNC decoders and MUD decoders are used to decode useful information from the overlapped signals transmitted simultaneously by different end users: (i) the PNC decoders attempt to decode for network-coded packets that are linear combinations of different users' native packets \referred{PNC06}\cite{PNC06}, while (ii) the MUD decoders attempt to decode for the individual native packets of the end users. At the MAC layer, the decoded PHY-layer packets from multiple time slots, including the network-coded and individual packets, are jointly used by a MAC layer decoder to decode for MAC-layer messages of the end users (see Section \ref{sec:overview} for details).

Previous work on NCMA \referred{NCMA1,NCMA2,MIMONCMA_Globecom}\cite{NCMA1,NCMA2,MIMONCMA_Globecom} focused on two end users only. Furthermore, both users adopted the same modulation for their signals. We refer to the previous NCMA as \emph{rate-identical} NCMA. This paper is the first attempt to generalize the prior two-user rate-identical NCMA system to a multi-user rate-diverse NCMA system. The motivations are as follows:

\begin{figure*}[t]
   \begin{minipage}{0.45\linewidth}
     \centering
     \includegraphics[width=0.65\textwidth]{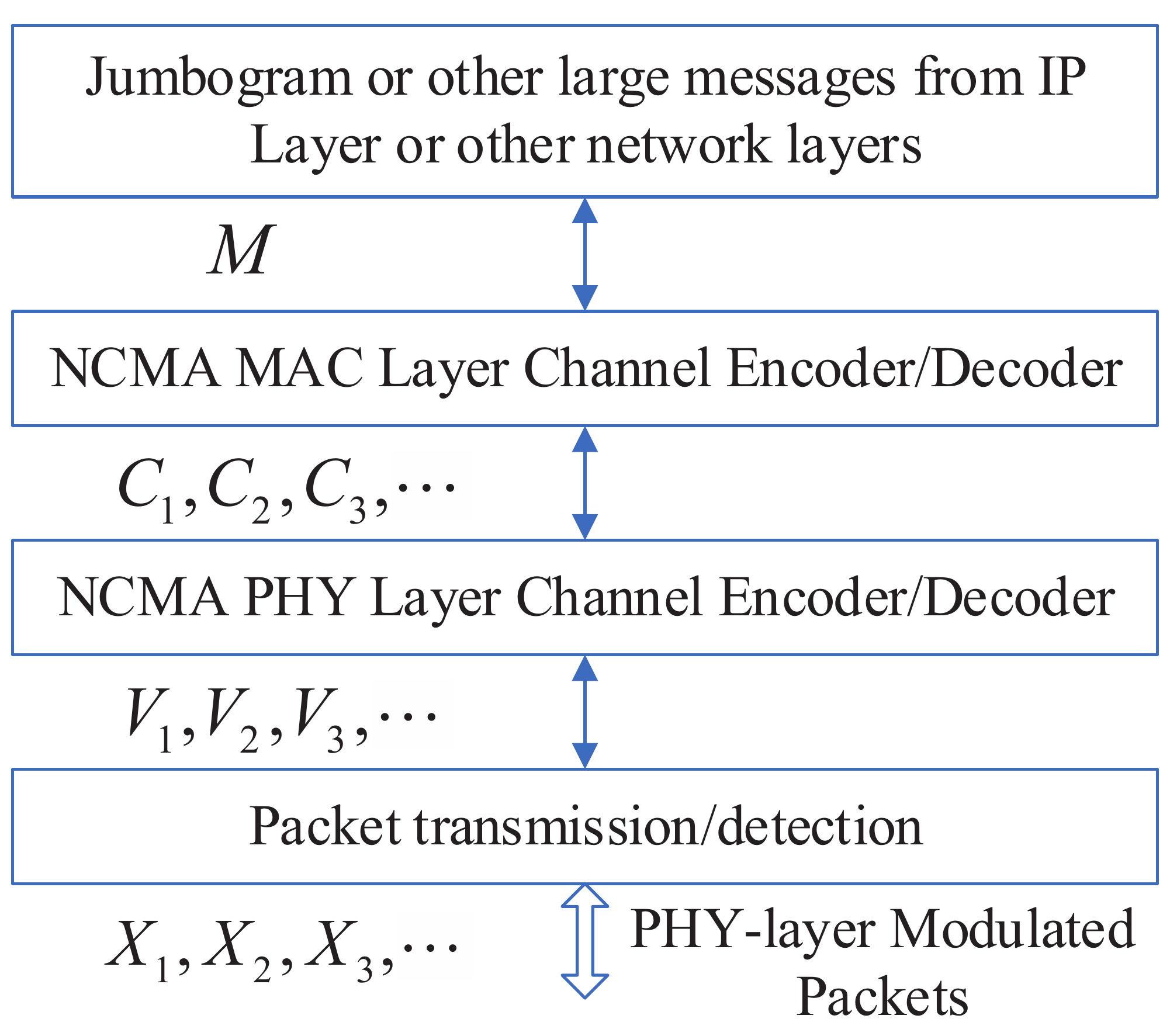}
     \caption{Encoding and decoding achitecture for NCMA end users.}\label{fig:general_architec}
   \end{minipage}
   \hfill
   \begin{minipage}{0.51\linewidth}
      \centering
      \includegraphics[width=1.1\textwidth]{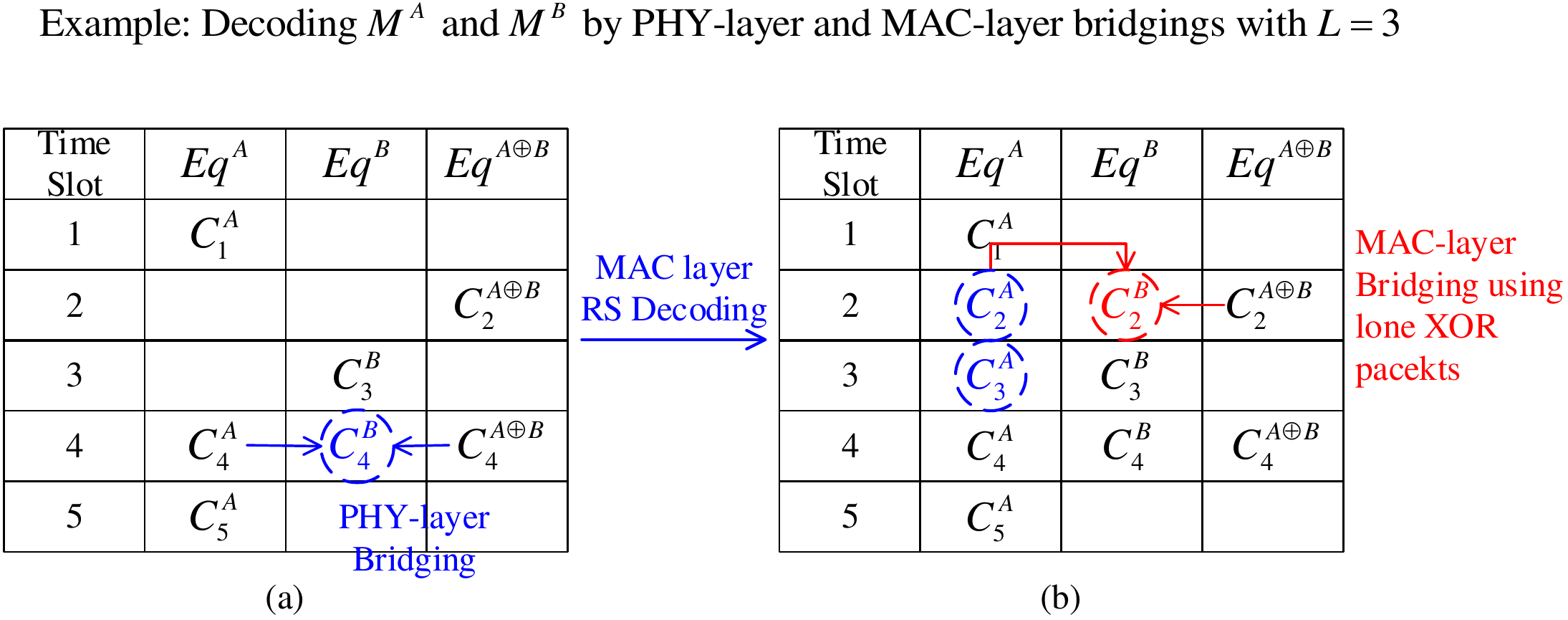}
      \caption{Two-user NCMA PHY-layer and MAC-layer bridgings example, using $L=3$ RS code: (a) PHY-layer bridging helps recover $C_4^B$ in time slot 4; (b) MAC-layer RS decoding and bridging for lone XOR packets.}\label{fig:two_user_example}
   \end{minipage}
\end{figure*}

\begin{itemize}\leftmargin=0in
\item [1)] In a multiuser random access network (e.g., ALOHA and CSMA networks), packet collisions from more than two users are inevitable if all users transmit with high probability (i.e., in a highly utilized network, we may encounter simultaneous transmissions by more than two users with high probability);
\item [2)] The uplink channels of different users to the AP may have different SNRs in practical scenarios. Forcing all users to use the same rate (modulation) may prevent the high-SNR users from fully exploiting their good channel conditions. As will be shown in Section \ref{sec:overview4}, this leads to low system throughput in rate-identical NCMA. In particular, the users with poor uplink channel conditions become the bottleneck of the whole system.
\end{itemize}

To better exploit different channel conditions and improve system throughputs, this paper considers the use of different modulations adopted by different users, referred to as \emph{rate-diverse} NCMA. We focus on a 3-user system to bring out the general issues in extending the rate-identical 2-user system to rate-diverse multi-user systems.

Although rate-diverse MUD has been widely studied, rate-diverse PNC decoder, an important component of the overall NCMA system, has not been investigated. In particular, rate-diverse PNC decoder for channel-coded PNC systems has not been well studied.  As will be shown in Section \ref{sec:ratediverse_ncma}, direct extension of the previous channel-coded PNC decoder, originally designed for homogeneous modulation among users, does not work when heterogeneous modulations are adopted, when the IEEE 802.11 channel coding scheme \referred{dot11std13}\cite{dot11std13} is adopted. We put forth a symbol-splitting channel coding and modulation scheme, referred to as \emph{symbol-splitting encoding}, to circumvent the issue encountered. We will show that symbol-splitting encoding enables channel-coded PNC decoding among different modulations, and by doing so allows the system to achieve substantially higher throughput than rate-identical channel-coded PNC systems.

To prove concept and to investigate system performance under real wireless environment, we implemented rate-diverse NCMA on software-defined radios. We implemented  a three-user NCMA system where two low-SNR users adopt BPSK, and one high-SNR user adopts QPSK. Our experiments show that with symbol-splitting encoding, rate-diverse NCMA outperforms rate-identical NCMA systems operated with BPSK-only and QPSK-only modulations by around 40\% and 80\%, respectively, in terms of total system throughput.



\section{Related Work}\label{sec:relatedwork}
\subsection{Physical-layer Network Coding (PNC)}\label{sec:relatedwork1}
PNC was originally proposed to increase the throughput of a two-way relay network (TWRN). It can double the throughput of a TWRN compared with the conventional store-and-forward relaying  \referred{PNC06}\cite{PNC06}. PNC has been studied and evaluated in depth during the past decade, and we refer the interested readers to \referred{liew2015primer,popovski2006anti,Nazer2011ReliablePNC}\cite{liew2015primer,popovski2006anti,Nazer2011ReliablePNC} and the references therein for details. Prior works on PNC focused almost exclusively on \emph{relay} networks. By contrast, NCMA was the first attempt to apply PNC in \emph{non-relay} networks (i.e., wireless multiple access networks) \referred{NCMA1,NCMA2,MIMONCMA_Globecom}\cite{NCMA1,NCMA2,MIMONCMA_Globecom}.

There have been some studies on PNC with different modulations in the literature. For example, \referred{HePNC,KoikeJSAC09}\cite{HePNC,KoikeJSAC09} considered non-channel-coded PNC schemes with different modulations. For reliable communication, channel-coded PNC is preferred \referred{liew2015primer}\cite{liew2015primer}. The schemes in \referred{HePNC,KoikeJSAC09}\cite{HePNC,KoikeJSAC09} are not applicable to channel-coded PNC because they do not preserve the linearity of the underlying channel codes. Our paper puts forth a symbol-splitting encoding scheme that preserves channel-code linearity when different users adopt different modulations, thereby enabling reliable rate-diverse PNC. As far as we know, this is the first rate-diverse channel-coded PNC design.

\subsection{Network Coding in Multiple Access Channels}\label{sec:relatedwork2}
Besides NCMA, recently there have been other efforts to apply network coding (including PNC) in multiple access networks. For example,  \referred{RAwithPNC,Cocco2011,SeekandDecode,CodedRandomAccess}\cite{RAwithPNC,Cocco2011,SeekandDecode,CodedRandomAccess} explored forming linear equations from the collided packets and derived source packets by solving the linear equations. However,  \referred{Cocco2011,CodedRandomAccess}\cite{Cocco2011,CodedRandomAccess} only compute one equation for each overlapped packet, whereas NCMA can have more than one equation for each overlapped packet under favourable channel condition. Furthermore, the decoding in \referred{RAwithPNC, SeekandDecode}\cite{RAwithPNC, SeekandDecode} is based on PHY-layer equations only, while NCMA makes use of an outer MAC-layer channel coding scheme to achieve better utilization of the  the PHY-layer PNC packets. Importantly, most existing works are theoretical in nature and lack implementation and experimental validations. They simply assume all users adopt the same signal modulations, even in fading channels. By contrast, our rate-diverse NCMA system takes into account the fact that different users are likely to experience different channel conditions under practical deployment scenarios.


\begin{figure}[t]
\centering
\includegraphics[width=0.48\textwidth]{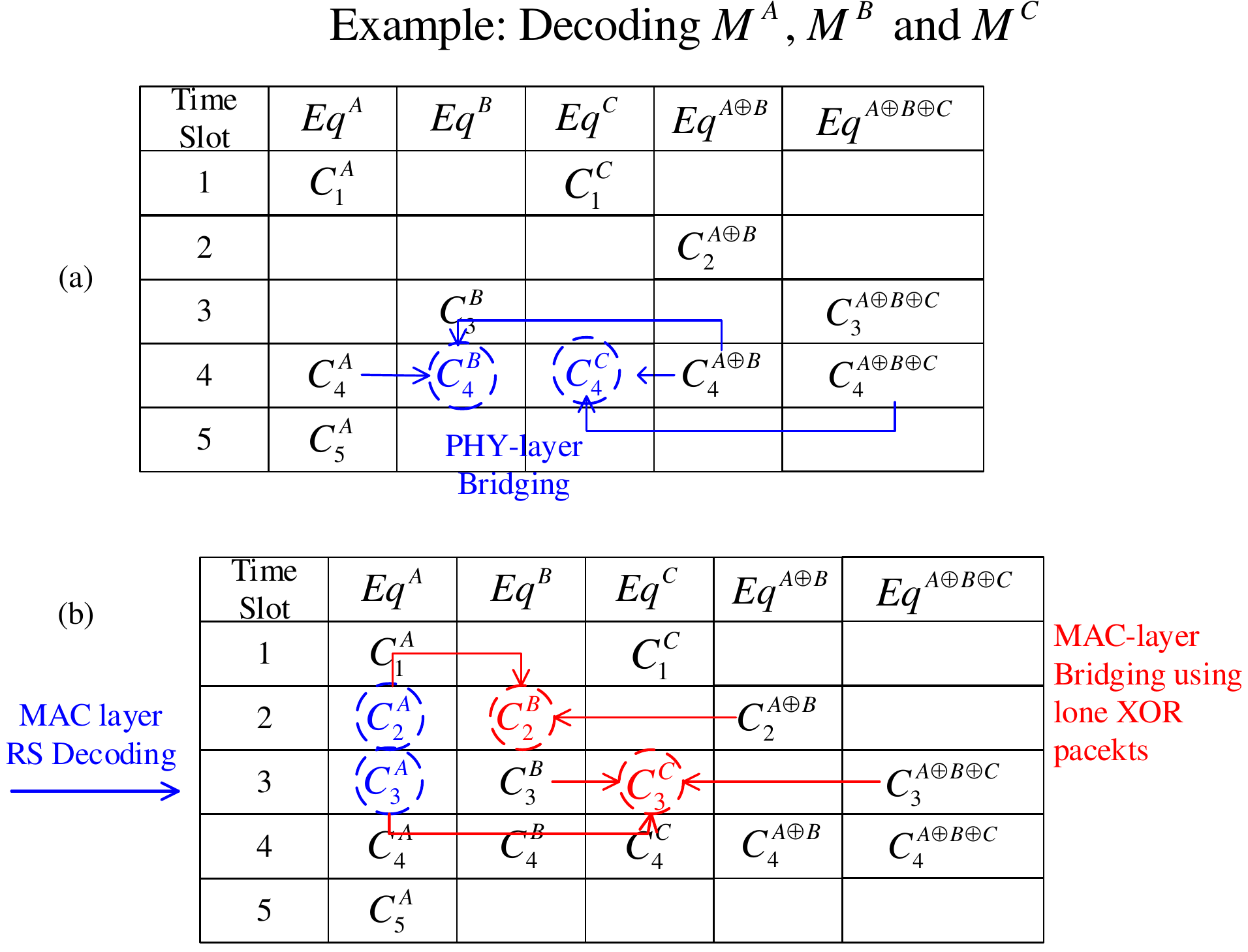}
\caption{Three-user NCMA PHY-layer and MAC-layer bridging example, generalized from Fig. \ref{fig:two_user_example}: (a) PHY-layer bridging; (b) MAC-layer RS decoding and bridging. }\label{fig:three_user_example}
\end{figure}

\section{Overview}\label{sec:overview}
\subsection{System Model}\label{sec:overview1}
We study a multiple access system where multiple end users transmit information to a common access point (AP) simultaneously (e.g., Fig. \ref{fig:system_model} shows an example with three end users). We consider the joint use of physical-layer network coding (PNC) and multiuser decoding (MUD) by the AP to boost system throughput. This system is referred to as a \emph{Network-Coded Multiple Access} (NCMA) system \referred{NCMA1}\cite{NCMA1}. In \referred{MIMONCMA_Globecom}\cite{MIMONCMA_Globecom}, the AP uses multiple antennas to accommodate high-order modulations beyond BPSK. The system is referred to as \emph{MIMO-NCMA}. In the rest of this paper, we focus on NCMA with two antennas at the AP, unless otherwise specified.

NCMA includes both MAC-layer and PHY-layer operations. With respect to Fig. \ref{fig:general_architec}, at the MAC layer, a large message $M^s$ of user $s$, $s \in \Theta  = \{ A,B,C,...\}$, is divided and encoded into multiple packets, $C_i^s$, $i=1,2,$... . We assume the use of Reed-Solomon (RS) code at the MAC layer when encoding a large message into multiple packets. At the PHY layer, each packet $C_i^s$ is further channel-encoded into $V_i^s$, and then modulated into $X_i^s$ for transmission. We adopt the standard IEEE 802.11 convolutional code as the PHY-layer channel codes. Throughout the whole paper, we focus on a time-slotted NCMA system\footnote{The general idea of NCMA can also be applied to carrier sense multiple access (CSMA) systems or time-division multiple access (TMDA) systems by modifying MAC protocols to allow simultaneous transmissions by users.}\referred{NCMA1}\cite{NCMA1}. In this system, each user $s$ transmits packets $X_1^s,X_2^s,...,X_i^s$  to the AP in successive time slots. Packets of different end users can be configured to be transmitted in the same time slot.

\subsection{Review of Two-user NCMA }\label{sec:overview2}
Let us briefly review the two-user NCMA system, and see how PNC and MUD can be jointly exploited to improve the system throughput. In the uplink phase (note: NCMA focuses on the uplink transmissions from end users to the AP), users A and B transmit simultaneously. The AP then decodes the superimposed signals using two multiuser decoders at the PHY layer: the MUD decoder and the PNC decoder. The MUD decoder attempts to decode both packets $C_i^A$ and $C_i^B$ explicitly, and the PNC decoder attempts to decode\footnote{This paper only considers the bit-wise eXclusive-OR (XOR) operation, $ \oplus $ ,  of $C_i^A$ and $C_i^B$.} $C_i^A \oplus C_i^B$. For each time slot $i$, a subset of $\{C_i^A,C_i^B,C_i^A \oplus C_i^B\}$  is successfully decoded. The successfully decoded PHY-layer packets in different times slots are then collected and passed to the MAC layer. With the MAC-layer RS code, the AP can recover the original messages $M^A$ and $M^B$ after collecting enough packets from the set ${\{ C_i^A,C_i^B,C_i^A \oplus C_i^B\} _{i = 1,2,...}}$.

We next illustrate the essence of NCMA with a simple example. Fig. \ref{fig:two_user_example}(a) shows an example of the decoding outcomes of the PNC and MUD decoder in five consecutive time slots. In time slot 4, $C_4^A$ and $C_4^A \oplus C_4^B$ (abbreviated as $C_4^{A \oplus B}$) are decoded. In this case, the PNC packet $C_4^{A \oplus B}$ can be used to recover the missing packet  $C_4^B$. This process, which leverages the complementary PNC XOR packet, is referred to as \emph{PHY-layer bridging} \referred{NCMA1}\cite{NCMA1}. However, PHY-layer bridging cannot be applied directly to time slot 2 because neither \emph{native} packet $C_2^A$  nor $C_2^B$ is available, and only the XOR packet $C_2^{A \oplus B}$  is decoded. In NCMA, such ``lone'' PNC packets, although not useful at the PHY layer, can be useful for MAC-layer decoding. In Fig. \ref{fig:two_user_example}(b), we assume the AP has recovered enough native packets $C_i^A$  to decode $M^A$ with the help of the MAC-layer RS code by time slot 5: in this example, $L=3$ PHY-layer packets are needed to recover $M^A$. This means that native packets $C_2^A$ and $C_3^A$ can also be recovered from $M^A$ (conceptually, we could re-encode $M^A$ to get $C_2^A$ and $C_3^A$, but in practice, more efficient procedure is available). Accordingly, the original ``lone" PNC packet $C_2^{A \oplus B}$ can now be combined with $C_2^A$ to recover $C_2^B$. Consequently, the AP also has enough native packets (i.e., $L=3$) to recover the message of B, $M^B$. We refer to this process as \emph{MAC-layer bridging}, which further boosts the system throughput by leveraging the ``lone'' PHY-layer PNC packets \referred{NCMA1}\cite{NCMA1}.

\subsection{Three-user NCMA}\label{sec:overview3}
Previous works on NCMA \referred{NCMA1,NCMA2,MIMONCMA_Globecom}\cite{NCMA1,NCMA2,MIMONCMA_Globecom}  were limited to the simple two-user case. In a wireless network with multiple users, if all users have a high probability to transmit, collisions with more than two users are inevitable. This paper investigates NCMA systems with three users as an example. In the following, we show that the underlying PHY-layer and MAC-layer decoder designs and bridging principles remain valid.

Suppose that users A, B, and C transmit their packets simultaneously in time slot $i$, and the AP receives their superimposed signals. At the PHY layer, three MUD decoders are needed for the AP to decode packets $C_i^A$, $C_i^B$ and $C_i^C$ individually. Similarly, four PNC decoders can be used to get the four network-coded combinations $C_i^A \oplus C_i^B$,  $C_i^A \oplus C_i^C$, $C_i^B \oplus C_i^C$ and $C_i^A \oplus C_i^B \oplus C_i^C$. That is, each PHY-layer decoder's output can be treated as a linear combination $aC_i^A \oplus bC_i^B \oplus cC_i^C$, where $a,b,c \in \{ 0,1\} $ and at least one of them must be 1. Fig. \ref{fig:three_user_example} shows an example of three-user NCMA by adding two more decoding outcome columns only, $C_i^C$ and $C_i^{A \oplus B \oplus C}$, to Fig. \ref{fig:two_user_example}.

It is worth emphasizing that there are more types of PHY-layer bridging for the three-user case than for the two-user case. For the three-user case, PHY-layer bridging can also happen between two PNC packets (namely, the XORed packets); for the two-user case, it can only happen between one PNC packet and one native packet. For instance, in time slot 4 of  Fig. \ref{fig:three_user_example}, the missing individual packet $C_4^C$ can be recovered by XORing $C_4^{A \oplus B}$ and $C_4^{A \oplus B \oplus C}$. Also, for the two-user case, all complementary PNC packets are resolved, and they do not need to be forwarded to the MAC layer. For the three-user case, a native packet and an ``unresolved'' packet can be forwarded to the MAC layer. For example, in time slot 3, the PNC packet $C_3^{A \oplus B \oplus C}$ is an ``unresolved'' packet even though the AP has obtained $C_3^B$ at the PHY layer, since no PHY-layer bridging happens between $C_3^B$ and $C_3^{A \oplus B \oplus C}$ (i.e., when we XOR $C_3^B$ with $C_3^{A \oplus B \oplus C}$, we have $C_3^{A \oplus C}$ that is a PNC packet rather than a native packet).

\subsection{Rate-Diverse NCMA\protect\footnote{Here, by ``rate", we mean the modulation order that corresponds to the PHY-layer data rate, assuming different users use the same baud rate and channel code. More generally, a rate-diverse system can also be created if different users use the same modulation, the same baud rate, but different channel codes at different code rates; however, in this case, XOR-CD decoding cannot be applied because of the use of different channel codes \referred{liew2015primer}\cite{liew2015primer}.}}\label{sec:overview4}
The previous work on NCMA assumes that all end users adopt the same modulation order, and we refer to this approach as \emph{rate-identical} NCMA. To support PHY-layer real-time processing, a non-iterative PNC decoder, called XOR-CD, is used in NCMA \referred{NCMA1,MIMONCMA_Globecom}\cite{NCMA1,MIMONCMA_Globecom}. XOR-CD first performs PNC decoding by finding the sample-wise XOR for each and every overlapping samples, and then performing channel decoding using the same channel decoder as that for the single-user system (i.e., the same channel decoder as that for A or B, assuming both A and B use the same channel code). The structure of XOR-CD decoder will be discussed in detail in Section \ref{sec:xorcd}. We remark that the XOR-CD decoder builds upon the same code rate and modulation order to maintain the linearity of the XORed channel-coded bits (i.e., for PNC packet $A \oplus B$, we have $\Pi \left( {A \oplus B} \right) = \Pi (A) \oplus \Pi (B)$, where $A$ and $B$ stand for user A and B's uncoded packets, and $\Pi \left(  \cdot  \right)$  stands for the channel encoding process).

\begin{figure}[t]
\centering
\includegraphics[width=0.48\textwidth]{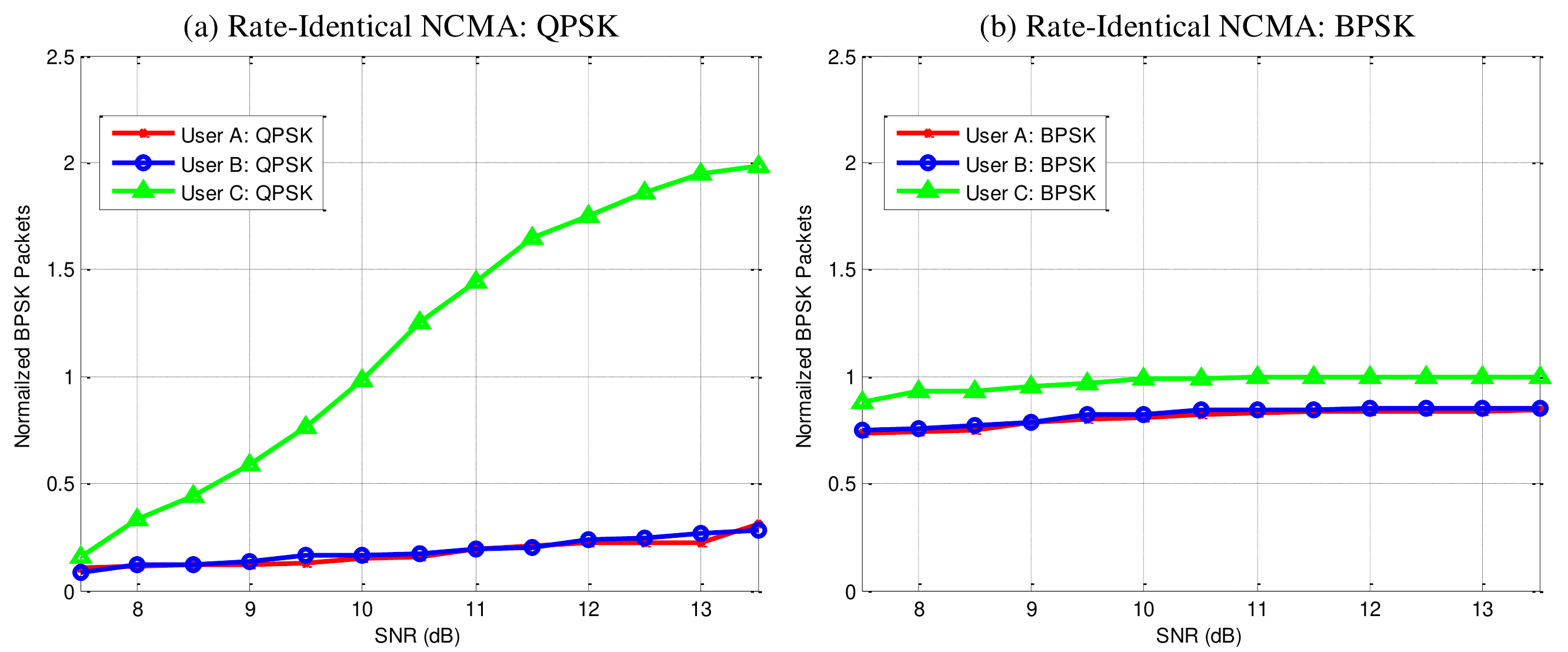}
\caption{Normalized throughputs of three-user rate-identical NCMA systems, assuming (a) QPSK and (b) BPSK. The y-axis stands for the normalized number of BPSK packets per time slot at the PHY layer, and the x-axis is the SNR of user C. Both users A and B's SNRs are set to be 7dB, and user C's SNR varies from 7.5dB to 15dB. }\label{fig:rate_identical_sim}
\end{figure}

Our latest investigation indicates that the rate-identical NCMA system faces a number of problems in the three-user system, especially when users have different channel conditions. Fig. \ref{fig:rate_identical_sim}(a) and Fig. \ref{fig:rate_identical_sim}(b) shows the throughputs per time slot of individual users in rate-identical NCMA systems generated by simulations, assuming QPSK and BPSK, respectively. In these simulations, users A and B's SNRs are equal (e.g., we fix $SNR_A$ = $SNR_B$ = 7dB), and user C's SNR varies from 7.5dB to 13.5dB. We assume AWGN channel, but the phase terms for different uplink channel gains are randomly chosen between 0 and $2\pi$; all users adopt the same rate-1/2 convolutional code as in the IEEE 802.11 standard. Also note that in Fig. \ref{fig:rate_identical_sim} we treat one QPSK packet as two BPSK packets for comparison. We observe the following:

\begin{itemize}\leftmargin=0in
\item [1)] In Fig. \ref{fig:rate_identical_sim}(a), all users adopt QPSK. Both users A and B have low throughputs because of their low SNRs, and the modulation order is not commensurate with the SNR, although user C's throughput approaches to 2 as its SNR increases;
\item [2)] In Fig. \ref{fig:rate_identical_sim}(b), all users adopt BPSK. Both users A and B can have higher throughputs than the QPSK case in Fig. \ref{fig:rate_identical_sim}(a). However, the throughput of user C is upper bounded by 1 and drops by around 100\% (i.e., from QPSK to BPSK), User C cannot leverage its higher SNR to obtain higher throughput because of its use of BPSK;
\item [3)] In both cases, the total system throughput is below 3.
\end{itemize}

For a practical multiple access system, it is unlikely that all users' uplink channel conditions at the AP are the same, since NCMA has no precoding (i.e., no power control and phase synchronization). Rate-identical NCMA forces all users to use the same modulation by ignoring their individual channel conditions (e.g., SNRs), and therefore the uplink with a poor channel condition becomes the bottleneck of the whole system. Here we ask a simple but fundamental question: \textbf{can NCMA allow different users to use different modulations; and if yes, how can the system throughput benefits by doing so? }

In the following, we present \emph{rate-diverse} NCMA, where different modulations are adopted by different users to better utilize the channel conditions. To accommodate rate diversity in NCMA, we have to address a critical issue: how can PNC mapping be performed under different modulations (e.g., the XOR operation between different modulated symbols), while maintaining the linearity of channel codes at the same time? To explain our approach, we first review the XOR-CD decoder in Section \ref{sec:xorcd}. Section \ref{sec:ratediverse_ncma} then presents our proposed channel coding and modulation designs.

\begin{figure}[t]
\centering
\includegraphics[width=0.49\textwidth]{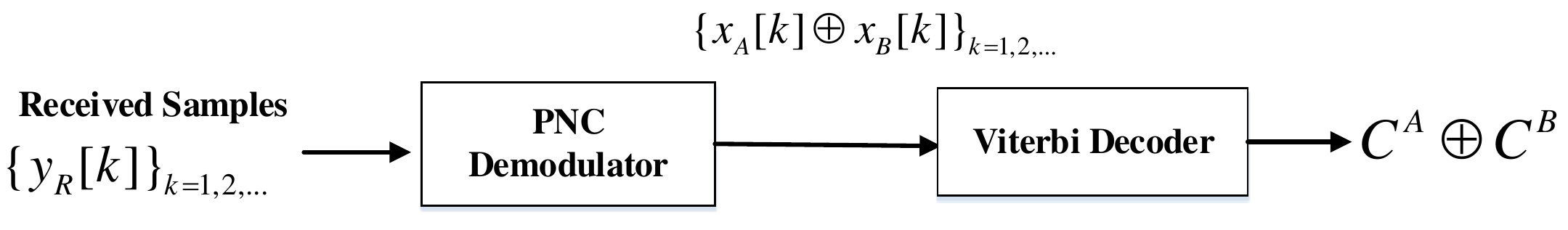}
\caption{NCMA PHY-layer PNC decoder (XOR-CD) Structure. }\label{fig:xorcd}
\end{figure}

\section{PNC Decoder Revisit: XOR-CD}\label{sec:xorcd}
This section revisits the XOR-CD decoder, an important building block of NCMA PHY-layer decoder. In this section, we explain that XOR-CD works when users adopt the same modulation order and code rate. The general architecture for XOR-CD is shown in Fig. \ref{fig:xorcd}. We adopt the $[133, 171]_8$ rate-1/2 convolutional code as in the IEEE 802.11 standard \referred{dot11std13}\cite{dot11std13}. A salient feature of XOR-CD is that the standard point-to-point Viterbi channel decoder can be used directly without any changes to support real-time decoding. For simplicity, here we assume the AP has one receive antenna, and two users, user A and user B, transmit packets   and   to the AP, respectively. Extensions to multiple antennas and multiple users can be found in \referred{NCMA2,MIMONCMA_Globecom}\cite{NCMA2,MIMONCMA_Globecom}.

\begin{figure}[t]
\centering
\includegraphics[width=0.45\textwidth]{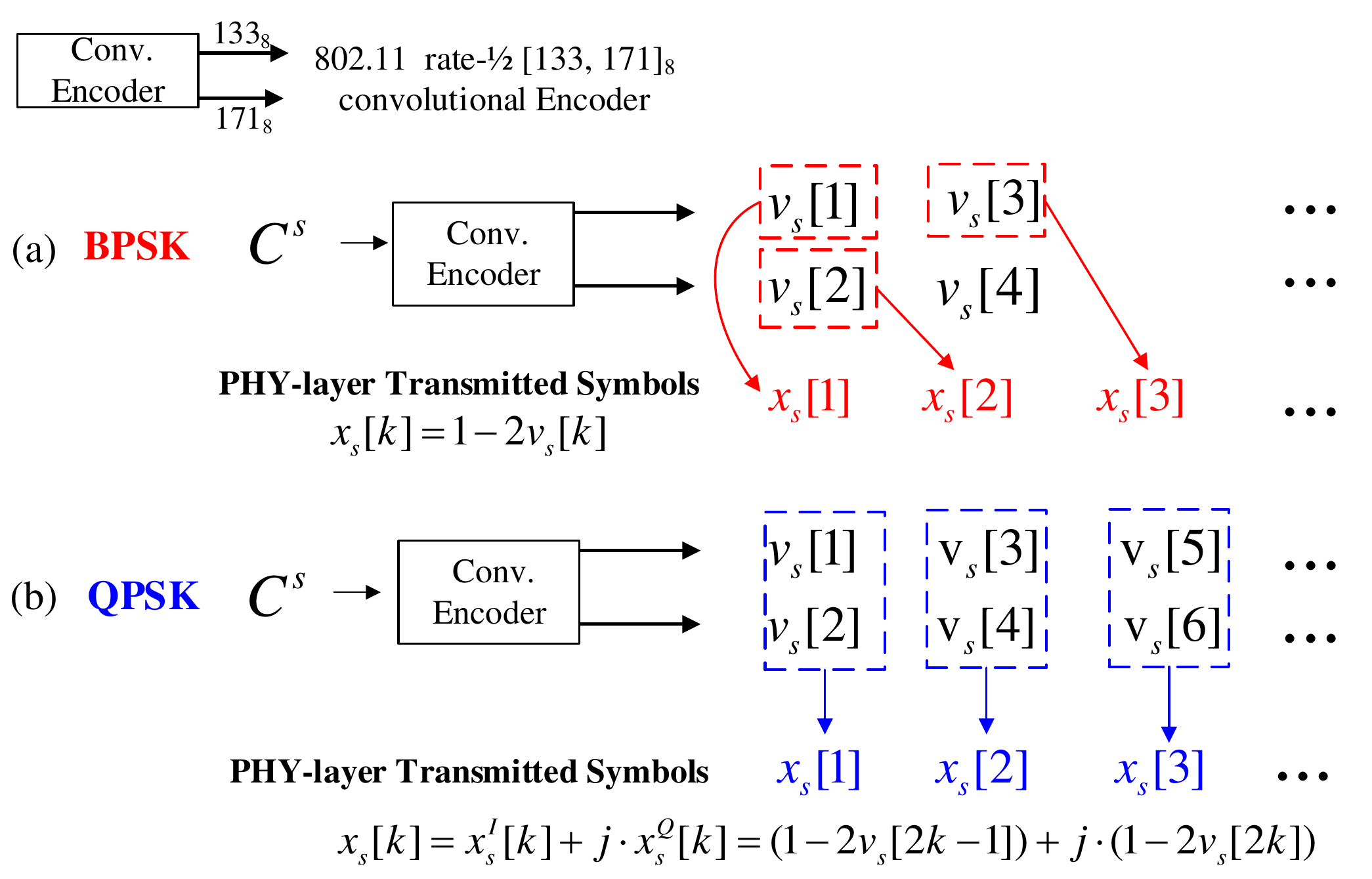}
\caption{Standard IEEE 802.11 convolutional encoding and modulation procedure under: (a) BPSK, and (b) QPSK modulations.}\label{fig:standard_coding}
\end{figure}

Let ${V^s} = ({v_s}[1],...,{v_s}[n],...)$ denote the PHY-layer codeword of user $s$ in one time slot (i.e., one binary convolutional-encoded packet of $C^s$), where ${V^s} = ({v_s}[1],...,{v_s}[n],...)$ is the n-th convolutional encoded bit. Assuming the modulation order is $m$ (e.g., $m =2$ for BPSK and $m =4$ for QPSK), the PHY-layer transmitted packet can be expressed as ${X^s} = ({x_s}[1],...,{x_s}[k],...)$, and ${x_s}[k]$ is the $k$-th modulated symbol of user $s$.

Let us assume an OFDM system where multipath fading can be dealt with by cyclic prefix (CP). The $k$-th received sample ${\{ {y_R}[k]\} _{k = 1,2,...}}$ in the frequency domain at the AP can be written as
\begin{align}
{y_R}[k] = {h_A}[k]{x_A}[k] + {h_B}[k]{x_B}[k] + w[k],
\label{equ:yR}
\end{align}

\noindent where $w[k]$ are white Gaussian noises (AWGN) with variances ${\sigma ^2}$, and ${h_s}[k]$ is the channel gain of the $k$-th sample of user $s$. In XOR-CD, the received samples ${\{ {y_R}[k]\} _{k = 1,2,...}}$ are first passed through a PNC demodulator to obtain the XOR bits ${\{ {v_A}[n] \oplus {v_B}[n]\} _{n = 1,2,...}}$. Note that the outputs from the PNC demodulator can be hard or soft bits. These XOR bits are then fed to a standard Viterbi decoder (as used in a point-to-point system) to decode the network-coded packet $C^A \oplus C^B$. Since the two users A and B make use of the same code rate, the standard Viterbi decoder can be used because XOR-CD exploits the linearity of linear channel codes (note: convolutional codes are linear; XOR-CD will work with other linear codes as well). Specifically, define $\Pi ( \cdot )$ as the channel coding operation. Since $\Pi ( \cdot )$ is linear, we have
\begin{align}
{V^A} \oplus {V^B} = \Pi \left( {{C^A}} \right) \oplus \Pi \left( {{C^B}} \right) = \Pi \left( {{C^A} \oplus {C^B}} \right).
\label{equ:linearity}
\end{align}

We now give an example on how to obtain ${\{ {v_A}[n] \oplus {v_B}[n]\} _{n = 1,2,...}}$, assuming BPSK modulation for both users. The $k$-th BPSK modulated symbol ${x_s}[k]$ of the PHY-layer transmitted packet ${X^s}$ can be expressed as ${x_s}[k] = 1 - 2{v_s}[k]$ (see Fig. \ref{fig:standard_coding}(a)). Let us focus on a particular set of symbols $\left( {{x_A}[k],{x_B}[k]} \right)$ from the two users. An important issue in PNC is how to calculate ${x_A}[k] \oplus {x_B}{\rm{[k]}}$ (abbreviated as ${x_{A \oplus B}}[k]$) using the received samples in (\ref{equ:yR}), referred to as \emph{PNC mapping}. The BPSK PNC mapping for ${x_{A \oplus B}}[k]$ is defined as
\begin{align}
{x_{A \oplus B}}[k] = {x_A}[k] \oplus {x_B}[k],
\label{equ:bpsk_mapping}
\end{align}

\noindent where ${x_A}[k] \oplus {x_B}[k]={x_A}[k]{x_B}[k]$  given that ${x_A}[k],{x_B}[k] \in \{ 1, - 1\} $. The demodulation rule for the XORed bits is defined as
\begin{align}
{v_A}[k] \oplus {v_B}[k] = \frac{{1 - {x_A}[k]{x_B}[k]}}{2}.
\label{equ:bpsk_demapping}
\end{align}

After that,  ${\{ {v_A}[n] \oplus {v_B}[n]\} _{n = 1,2,...}}$ are fed to the Viterbi decoder to decode the PNC packet ${C^A} \oplus {C^B}$. When different users adopt the same modulation, XOR-CD works in a similar way for higher-order modulations beyond BPSK (e.g., QPSK, 16-QAM) after each modulated symbol is mapped to bits \referred{MIMONCMA_Globecom}\cite{MIMONCMA_Globecom}. However, we will show in the next section that XOR-CD cannot be used directly when users use different modulations.

\section{Rate-Diverse NCMA System}\label{sec:ratediverse_ncma}
This section presents rate-diverse NCMA. We study the case of three-user NCMA with two users, say users A and B, adopting BPSK, and one user, say user C, adopting QPSK (abbreviated as 2B1Q). We remark that the decoding principle for 2B1Q can be generalized to other scenarios easily, e.g., 2QPSK+1BPSK.

For 2B1Q, PNC decoding (XOR-CD) between the two BPSK users is the same as before (e.g., we can adopt the BPSK PNC mapping defined in (\ref{equ:linearity}), and the calculations of each XORed bit's soft information will be presented in Section \ref{sec:ratediverse_ncma3}). However, we show in Section \ref{sec:ratediverse_ncma1} that conventional XOR-CD does not work between BPSK and QPSK users. Section \ref{sec:ratediverse_ncma2} then presents our designs to enable PNC even among different modulations. Section \ref{sec:ratediverse_ncma3} presents the details of our rate-diverse NCMA PHY-layer decoders.

\subsection{Problem of PNC Decoder with Different Modulations }\label{sec:ratediverse_ncma1}
We first explain the problem of XOR-CD with different modulations. Let us focus on the set of symbols $\left( {{x_A}[k],{x_B}[k],{x_C}[k]} \right)$ from the three users, where ${x_A}[k]$ and ${x_B}[k]$ are BPSK modulated symbols as in Fig. \ref{fig:standard_coding}(a). Assuming QPSK modulation for user C, the $k$-th modulated symbol ${x_C}[k]$ of the PHY-layer transmitted packet ${X^C}$ is
\begin{align}
{x_C}[k] = (1 - 2{v_C}[2k - 1]) + & j \cdot (1 - 2{v_C}[2k]),\notag \\
&{\rm{       }}k = 1,2,...,n,...
\label{equ:qpsk_standard}
\end{align}

\noindent That is, the odd (even) bits of the convolutional-encoded packet $V^C$ are mapped to the in-phase (quadrature) component of ${x_C}[k]$ in QPSK, i.e., $x_C^I[k] = 1 - 2{v_C}[2k - 1]$ ($x_C^Q[k] = 1 - 2{v_C}[2k]$). Fig. \ref{fig:standard_coding} illustrates the differences between BPSK and QPSK modulations. Note that for both BPSK and QPSK, the odd bits and even bits of $V^s$ are generated from two different code generator polynomials (i.e., $133_8$ and $171_8$ in the IEEE 802.11 Standard). Since each QPSK symbol contains two bits, one from each polynomial, while each BPSK symbol contains only one bit from one of the polynomials, how to perform the proper PNC mapping (XOR-CD) for the overlapping QPSK and BPSK symbols is an issue. It is difficult to find a proper PNC mapping between a BPSK symbol and a QPSK symbol (e.g., ${x_A}[k] \oplus {x_C}[k]$ and ${x_B}[k] \oplus {x_C}[k]$) that maintains the linearity of convolutional codes as (\ref{equ:linearity}).

\begin{figure}[t]
\centering
\includegraphics[width=0.49\textwidth]{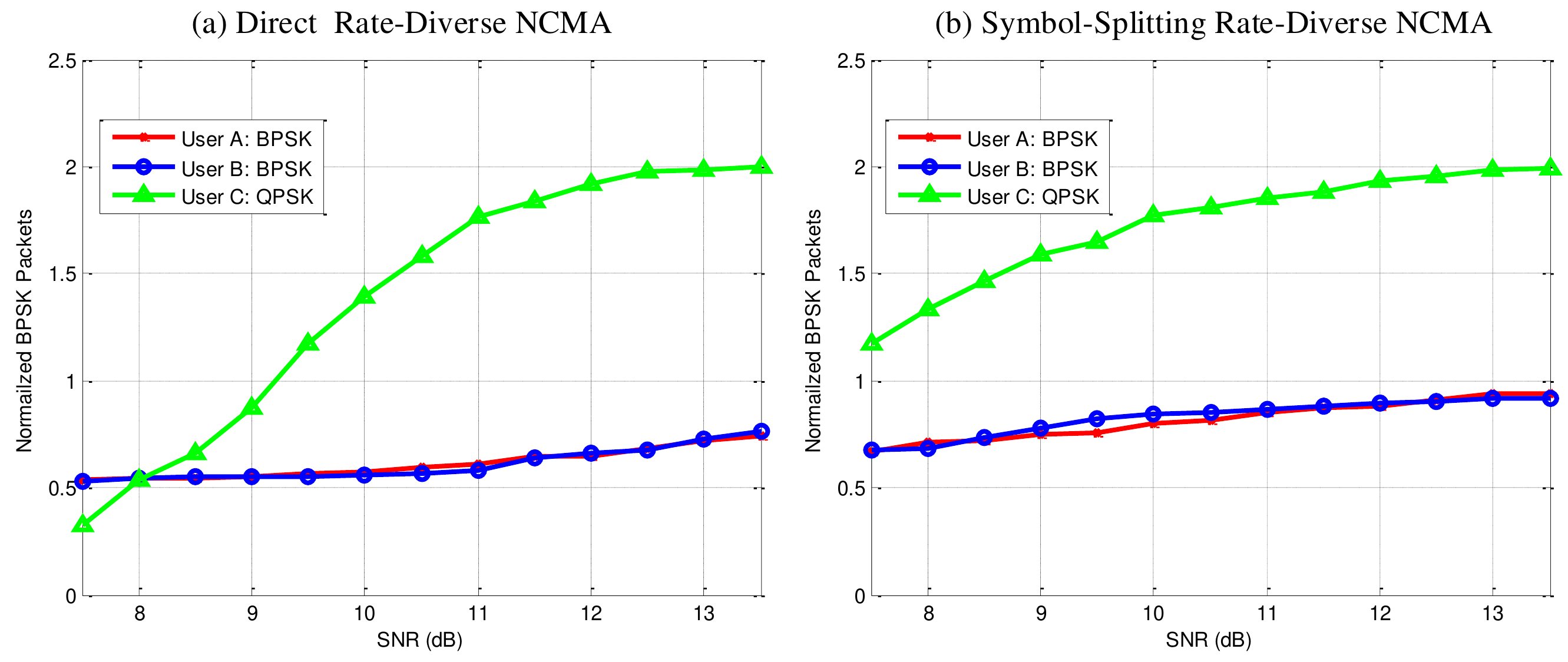}
\caption{Normalized throughputs of three-user Rate-Diverse NCMA systems: (a) Direct Rate-Diverse NCMA and (b) Symbol-Splitting Rate-Diverse NCMA. We assume users A and B adopt BPSK, and user C adopts QPSK. The y-axis stands for the normalized number of BPSK packets per time slot at the PHY layer, and the x-axis is the SNR of user C. Both users A and B's SNRs are set to be 7dB, and user C's SNR varies from 7.5dB to 15dB. }\label{fig:rate_diverse_sim}
\end{figure}

As a result, if we directly generalize rate-identical NCMA to rate-diverse NCMA, only one possible PNC decoder is available (i.e., to decode $C^A \oplus C^B$). We refer to such an NCMA system as \emph{Direct Rate-Diverse} NCMA (DR-NCMA). Fig. \ref{fig:rate_diverse_sim}(a) shows the throughputs of individual users of DR-NCMA with the same simulation setups as Fig. \ref{fig:rate_identical_sim}, except that users A and B adopt BPSK, and user C adopts QPSK. Compared with BPSK rate-identical NCMA in Fig. \ref{fig:rate_identical_sim}(b), the BPSK users have lower throughputs in DR-NCMA. Moreover, the performance of the QPSK user is limited by the MUD decoder (e.g., no PHY-layer and MAC-layer bridgings for the QPSK user). Overall, DR-NCMA does not fully exploit the advantages of NCMA, although the total system throughput is now above 3 (when user C has high SNRs). Fortunately, as will be seen in the next subsection, we can redesign the channel coding and modulation scheme that can enable PNC among different modulations.

\subsection{Exploit PNC between Different Modulations  }\label{sec:ratediverse_ncma2}
We now present our designs that enable PNC among different modulations. Let us focus on user A (BPSK) and user C (QPSK) as an example. We study how to perform PNC mapping between ${x_A}[k]$ and ${x_C}[k]$, where ${x_C}[k] = x_C^I[k] + j \cdot x_C^Q[k]$.

As discussed in Section \ref{sec:ratediverse_ncma1}, an important design issue for PNC mapping is how to maintain the linearity of convolutional codes. Conventional channel encoding and modulation scheme fails to do so because within the overlapping QPSK and BPSK symbols, $x_C^I[k]$ and $x_C^Q[k]$ are from two different polynomials, while ${x_A}[k]$ is from one of the two polynomials. However, if the in-phase and quadrature bits of the QPSK packet can be encoded separately like two BPSK packets (one containing in-phase bits; one containing quadrature bits), in which alternate bits of the in-phase and quadrature bits are also taken from the two different polynomials, then $x_A[k]$, $x_C^I[k]$ and $x_C^Q[k]$ can come from the same polynomial for the same $k$ (i.e., either $133_8$ or $171_8$ in the IEEE standard). PNC mappings that maintain the linearity of convolutional codes become possible among $x_A[k]$, $x_C^I[k]$ and $x_C^Q[k]$.

\begin{figure}[t]
\centering
\includegraphics[width=0.45\textwidth]{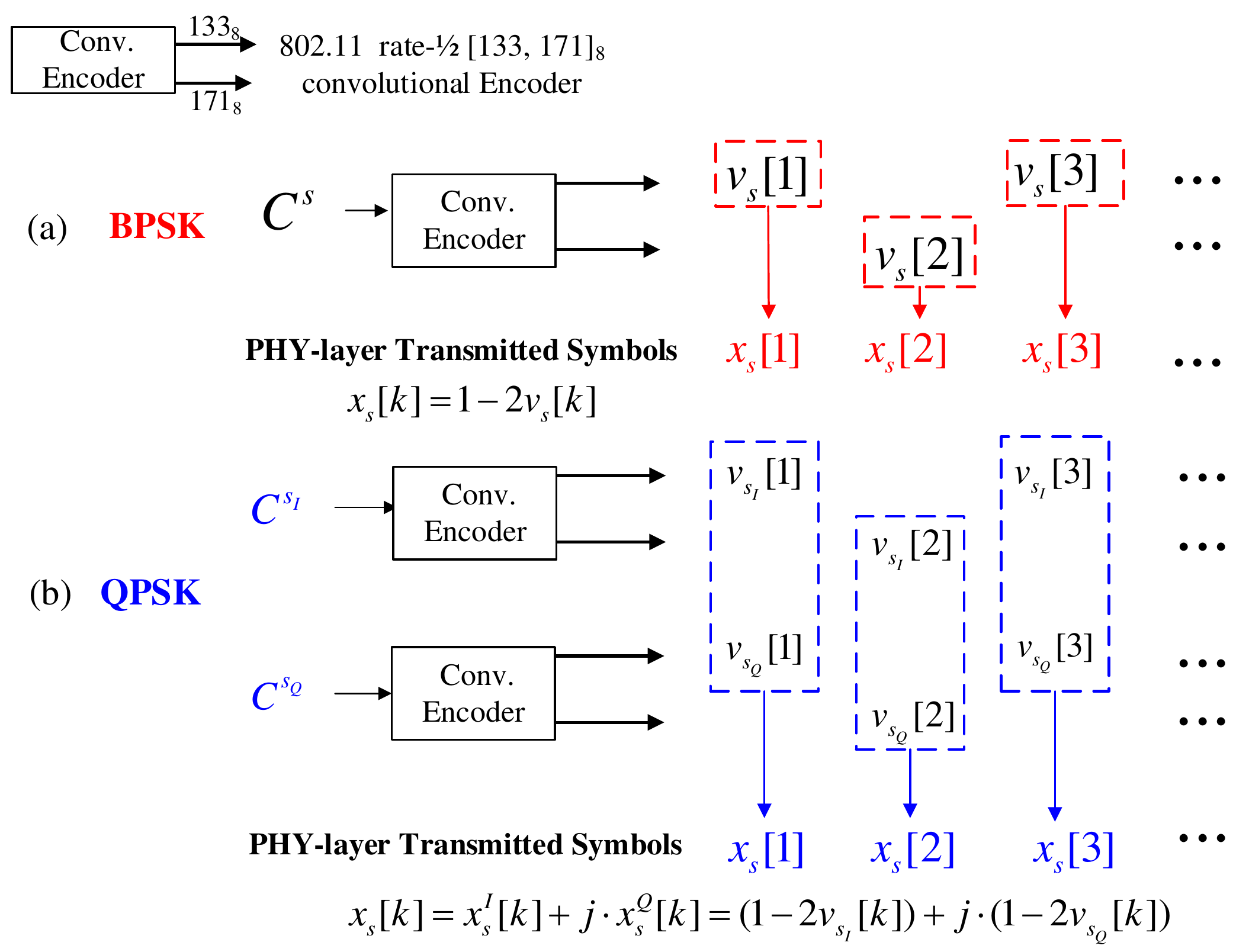}
\caption{Convolutional encoding and modulation schemes for Symbol-Splitting Rate-Diverse NCMA: (a) Same procedure as in the IEEE 802.11 standard for BPSK packets (the same figure as Fig. \ref{fig:standard_coding}, and (b) Symbol-Splitting Encoding for QPSK packets.}\label{fig:sp_coding}
\vspace{-0.15in}
\end{figure}

Fig. \ref{fig:sp_coding} presents our channel encoding and modulation scheme for QPSK in rate-diverse NCMA. For user C, let $C^{{C_I}}$ and $C^{{C_Q}}$ denote two small packets (which can be equally divided from $C^{C}$). They are separately convolutional encoded to ${V^{{C_I}}} = \{ {v_{{C_I}}}[1],{v_{{C_I}}}[2],...,{v_{{C_I}}}[n],...\}$ and ${V^{{C_Q}}} = \{ {v_{{C_Q}}}[1],{v_{{C_Q}}}[2],...,{v_{{C_Q}}}[n],...\}$, respectively. The $k$-th modulated symbol  ${x_C}[k]$ for the QPSK packet ${X^C} = ({x_C}[1],...,{x_C}[k],...)$ is now defined as
\begin{align}
{x_C}[k] &= x_C^I[k] + j \cdot x_C^Q[k] \notag \\
&= (1 - 2{v_{{C_I}}}[k]) + j \cdot (1 - 2{v_{{C_Q}}}[k]).
\label{equ:qpsk_qp}
\end{align}
\noindent That is, $C^{{C_I}}$ ($C^{{C_Q}}$) is encoded to be the in-phase (quadrature) bits of the QPSK packet ${X^C}$. We refer to this channel encoding and modulation scheme as \emph{symbol-splitting encoding}.

In essence, the symbol-splitting encoding scheme makes one QPSK packet equivalent to two ``small'' BPSK packets from the channel coding perspective, e.g., two BPSK packets are embedded in the in-phase and quadrature parts of one QPSK packet, respectively. Since each ``small'' BPSK packet is now encoded in the same way as a regular BPSK packet, we can define the PNC mapping between symbols ${x_A}[k]$ and ${x_C}[k]$ as\footnote{We can also define PNC mapping $x_C^I[k] \oplus x_C^Q[k]$ and compute the PNC packet ${C^{{C_I}}} \oplus C^{{C_Q}}$.  In symbol-splitting encoding, ${C^{{C_I}}}$ and ${C^{{C_Q}}}$ can be regarded as two packets with a fixed 90-degree relative phase offset (i.e., they will be encoded as the in-phase and quadrature parts of the QPSK packet). However, our simulation results show that ${C^{{C_I}}} \oplus C^{{C_Q}}$ does not give extra performance gain since MUD decoders that decode ${C^{{C_I}}}$ and ${C^{{C_Q}}}$ work well already. In this paper, we do not consider the PNC decoders that contain ${C^{{C_I}}} \oplus C^{{C_Q}}$. }
\begin{align}
x_{A \oplus C}^I[k] = {x_A}[k] \oplus x_C^I[k], \notag \\
x_{A \oplus C}^Q[k] = {x_A}[k] \oplus x_C^Q[k].
\label{equ:rate_diverse_mapping}
\end{align}

With the same demodulation rule as in (\ref{equ:bpsk_demapping}), the corresponding XOR bits ${\{ {v_A}[n] \oplus {v_{{C_I}}}[n]\} _{n = 1,2,...}}$ and ${\{ {v_A}[n] \oplus {v_{{C_Q}}}[n]\} _{n = 1,2,...}}$ obtained from the demodulator are then fed to the Viterbi decoder to decode ${C^A} \oplus C^{{C_I}}$ and ${C^A} \oplus C^{{C_Q}}$, respectively. That is, with symbol-splitting encoding, we can perform PNC decoding between BPSK and QPSK users.

\subsection{Symbol-Splitting Rate-Diverse NCMA }\label{sec:ratediverse_ncma3}
This section presents the rate-diverse NCMA system with symbol-splitting encoding scheme for QPSK packet. We refer to this NCMA system as \emph{Symbol-splitting Rate-Diverse} NCMA (SR-NCMA). We first list down different PHY-layer decoders used in SR-NCMA, and present the SR-NCMA system throughput compared with DR-NCMA and rate-identical NCMA. After that, we present the details of PHY-layer decoders for SR-NCMA.

Section \ref{sec:ratediverse_ncma2} discussed two PNC decoders that decode ${C^A} \oplus C^{{C_I}}$ and ${C^A} \oplus C^{{C_Q}}$ between the BPSK user A and the QPSK user C. In general, with symbol-splitting encoding, there are total seven possible PNC decoders to decode different linear combinations between the three users A, B, and C, as shown in Table \ref{tab:rate_diverse_decoder}. Also, four MUD decoders can be used in SR-NCMA. In short, each PHY-layer decoder's output can be treated as a linear combination\footnote{The PHY-layer decoding complexity for SR-NCMA is acceptable. At the PHY layer, SR-NCMA and DR-NCMA amount to decode 11 and 5 equivalent BPSK packets (including MUD and PNC packets, and one QPSK packet is treated as two BPSK packets). The QPSK and BPSK rate-identical NCMA amount to decode 14 and 7 BPSK packets, respectively. As will be seen in the Section \ref{sec:Exp2}, the total system throughput of SR-NCMA can be up to 80\% higher than that of DR-NCMA and rate-identical NCMA.} $aC^A \oplus bC^B \oplus cC^{{C_I}}$ or $aC^A \oplus bC^B \oplus cC^{{C_Q}}$ , where $a,b,c \in \{ 0,1\} $ and at least one of them must be 1.

Fig. \ref{fig:rate_diverse_sim}(b) shows the throughputs of individual users of SR-NCMA. Compared with DR-NCMA in Fig. \ref{fig:rate_diverse_sim}(a), the BPSK users have higher throughputs in SR-NCMA, which is also comparable to BPSK rate-identical NCMA in Fig. \ref{fig:rate_identical_sim}(b). The throughput of the QPSK user also improves and converges to 2 quickly as SNR increases, thanks to the PNC packets between QPSK and BPSK users, e.g., the QPSK user can have PHY-layer and MAC-layer bridgings through PNC packets in SR-NCMA. For the total system throughput, SR-NCMA has the highest throughput compared with DR-NCMA and rate-identical NCMA (e.g., approaches to 4 when user C has high SNRs). Overall, SR-NCMA allows users to select a proper modulation order to better utilize the channel conditions.

\noindent \textbf{Calculations of Soft Information in Demodulators}

We now explain how to obtain the soft information from the PHY-layer demodulators. We focus on the soft information of ${\{ {v_A}[n] \oplus {v_{{C_I}}}[n]\} _{n = 1,2,...}}$ as an example, and other PNC and MUD demodulators can follow the same manner. The soft information of ${\{ {v_A}[n] \oplus {v_{{C_I}}}[n]\} _{n = 1,2,...}}$ are fed to the Viterbi decoder to decode packet ${C^A} \oplus C^{{C_I}}$.

Let the received frequency-domain samples on the two antennas at the AP be ${{\rm{\{ }}{y_{R1}}[k]{\rm{\} }}_{k = 1,2,3...}}$ and ${{\rm{\{ }}{y_{R2}}[k]{\rm{\} }}_{k = 1,2,3...}}$ (our NCMA system is an OFDM system). Our target is to compute the log-likelihood ratio (LLR) of ${v_A}[k] \oplus {v_{{C_I}}}[k]$, based on the $k$-th received samples ${y_{R1}}[k]$ and ${y_{R2}}[k]$ (in the following, we drop the index k for simplicity):
\begin{align}
{y_{R1}} = {h_{A1}}{x_A} + {h_{B1}}{x_B} + {h_{C1}}{x_C} + {w_1}, \notag \\
{y_{R2}} = {h_{A2}}{x_A} + {h_{B2}}{x_B} + {h_{C2}}{x_C} + {w_2}.
\label{equ:y1y2}
\end{align}

\noindent where ${h_{s1}}$ and ${h_{s2}}$ are the uplink channel gains of end user s associated with the first and second antenna, respectively, and ${w_1}$ and ${w_2}$ are additive white Gaussian noises (AWGN) with variances ${\sigma _1}^2$ and ${\sigma _2}^2$. We assume the noise variances ${\sigma _1}^2$ and ${\sigma _2}^2$ to be the same. Note that, in real wireless systems, ${\sigma _1}^2$ and ${\sigma _2}^2$ may not be equal sometimes; however, our derivations below can be easily generalized to deal with the case $\sigma _1^2 \ne \sigma _2^2$.

Define the LLR of packet A's BPSK symbol (i.e., $x_A$) as $log(P_A/Q_A)$, where $P_A$ and $Q_A$ are the probabilities for $x_A$ to be 1 and -1, respectively. Similarly, for $LLR(x_A \oplus x_C^I)$,  $P_{A \oplus {C_I}}$ and $Q_{A \oplus {C_I}}$ are the probabilities corresponding to $x_A \oplus x_C^I = 1$ and $x_A \oplus x_C^I = -1$. We have
\begin{align}
LLR(x_A \oplus x_C^I) &= \log P_{A \oplus {C_I}}- \log Q_{A \oplus {C_I}} \notag\\
&= \log \Pr (x_A \oplus x_C^I = 1|{y_{R1}},{y_{R2}}) \notag\\
&- \log \Pr (x_A\oplus x_C^I =  - 1|{y_{R1}},{y_{R2}}).
\label{equ:llr}
\end{align}

\newcounter{mytempeqncnt}
\begin{figure*}[!t]
\setcounter{mytempeqncnt}{\value{equation}}
\small
\begin{align}
\log &P_{A \oplus {C_I}}= \log \Pr (x_A \oplus x_C^I = 1|{y_{R1}},{y_{R2}}) \notag\\
&{\rm{     }} \propto \log \sum\limits_{({x_A},{x_B},{x_C}) \in {\chi _{{x_{A \oplus {C_I}}} = 1}}} {\exp \{  - \frac{{|{y_{R1}} - {h_{A1}}{x_A} - {h_{B1}}{x_B} - {h_{C1}}{x_C}{|^2}}}{{\sigma^2}}} {\rm{\} exp\{ }} - \frac{{|{y_{R2}} - {h_{A2}}{x_A} - {h_{B2}}{x_B} - {h_{C2}}{x_C}{|^2}}}{{\sigma^2}}{\rm{\} }}
\label{equ:logpnc_p1}
\end{align}
\hrulefill
\vspace*{-10pt}
\end{figure*}

\begin{figure*}[!t]
\setcounter{mytempeqncnt}{\value{equation}}
\small
\begin{align}
\log P_{A \oplus {C_I}}^{} &\propto {\max _{({x_A},{x_B},{x_C}) \in {\chi _{{x_{A \oplus {C_I}}} = 1}}}}\{  - |{y_{R1}} - {h_{A1}}{x_A} - {h_{B1}}{x_B} - {h_{C1}}{x_C}{|^2} - |{y_{R2}} - {h_{A2}}{x_A} - {h_{B2}}{x_B} - {h_{C2}}{x_C}{|^2}{\rm{\} }} \notag\\
&{\rm{         }} \propto {\min _{({x_A},{x_B},{x_C}) \in {\chi _{{x_{A \oplus {C_I}}} = 1}}}}\{ |{y_{R1}} - {h_{A1}}{x_A} - {h_{B1}}{x_B} - {h_{C1}}{x_C}{|^2} + |{y_{R2}} - {h_{A2}}{x_A} - {h_{B2}}{x_B} - {h_{C2}}{x_C}{|^2}{\rm{\} }}
\label{equ:logpnc_p1_sim}
\end{align}
\hrulefill
\vspace*{-10pt}
\end{figure*}

\noindent Out of the 16 constellation points associated with the symbols $({x_A},{x_B},{x_C})$, let ${\chi _{{x_{A \oplus {C_I}}} = 1}}$ denote the set of symbols $({x_A},{x_B},{x_C})$ that satisfy $x_A \oplus x_C^I = 1$.  We can express $\log P_{A \oplus {C_I}}$ as (\ref{equ:logpnc_p1}). $\log Q_{A \oplus {C_I}}$ can be computed in a similar way based on the set ${\chi _{{x_{A \oplus {C_I}}} =  - 1}}$. To further simplify (\ref{equ:logpnc_p1}), we adopt the log-max approximation, $\log (\sum {_ie} xp({z_i})) \approx {\max _i}{z_i}$. For example, $\log P_{A \oplus {C_I}}$ can be expressed as (\ref{equ:logpnc_p1_sim}).

Note that after simplification, the AP does not need to estimate the noise variance $\sigma^2$ in (\ref{equ:logpnc_p1_sim}). The physical meaning of (\ref{equ:logpnc_p1_sim}) can be understood to be selecting one constellation point with the minimum Euclidean distance among all symbols $({x_A},{x_B},{x_C})$ in set ${\chi _{{x_{A \oplus {C_I}}} = 1}}$ for computing $\log P_{A \oplus {C_I}}$ (similarly, select one constellation point in ${\chi _{{x_{A \oplus {C_I}}} =-1}}$ for computing $\log Q_{A \oplus {C_I}}$). After that, we substitute $\log P_{A \oplus {C_I}}$ and $\log Q_{A \oplus {C_I}}$ into (\ref{equ:llr}) to obtain the LLR. The demodulation from $x_A[k] \oplus x_C^I[k]$ to ${v_A}[k] \oplus v_{{C_I}}[k]$ is a one-to-one mapping (see (\ref{equ:bpsk_demapping})), and the following LLR relationship always holds
\begin{align}
LLR({v_A}[k] \oplus v_{{C_I}}[k]){\rm{ = }}LLR(x_A[k] \oplus x_C^I[k]).
\label{equ:pnc_demap}
\end{align}

\begin{table}[t]
\centering
\caption{\textnormal{SR-NCMA PHY-layer Decoders, assuming three end users A, B and C. Users A and B use BPSK, and user C uses QPSK.}}
\begin{tabular}{|c|c|c|c|c|}
 \hline
 {MUD Decoder} & \multicolumn{2}{c|}{PNC Decoder} \\
 \hline
 ${C^A}$        & ${C^A} \oplus {C^B}$        & \\  \hline
 ${C^B}$         & ${C^A} \oplus C^{{C_I}}$       & ${C^A} \oplus C^{{C_Q}}$  \\  \hline
 $C^{{C_I}} $ & ${C^B} \oplus C^{{C_I}}$ &  ${C^B} \oplus C^{{C_Q}}$\\  \hline
 $C^{{C_Q}}$ &  ${C^A} \oplus {C^B} \oplus C^{{C_I}}$ & ${C^A} \oplus {C^B} \oplus C^{{C_Q}}$ \\  \hline

 \end{tabular}
 \label{tab:rate_diverse_decoder}
\end{table}

\section{Experimental Results} \label{sec:Exp}
To evaluate the performance of our symbol-splitting rate-diverse NCMA system, we implemented it on software-defined radios. Section \ref{sec:Exp1} presents the implementation details and experimental setup, and Section \ref{sec:Exp2} presents the experimental results.

\subsection{Implementation Details and Experimental Setup} \label{sec:Exp1}
The rate-diverse NCMA system was built on the USRP hardware \referred{Ettus}\cite{Ettus} and the GNU Radio software with the UHD hardware driver. We extended the rate-identical NCMA system in \referred{MIMONCMA_Globecom}\cite{MIMONCMA_Globecom} as follows:
\begin{itemize}\leftmargin=0in
\item [a)] We modified the transceiver design in \referred{MIMONCMA_Globecom}\cite{MIMONCMA_Globecom} so as to support three users in addition to two users;
\item [b)] We modified the conventional convolutional encoding and modulation scheme in the 802.11 standard to the symbol-splitting encoding for QPSK packets so as to enable PNC decoding among different modulations, as discussed in Section \ref{sec:ratediverse_ncma2};
\item [c)] We implemented the XOR-CD and MUD-CD decoders in symbol-splitting rate-diverse NCMA as described in Section \ref{sec:ratediverse_ncma3}.
\end{itemize}

For experimentation, we deployed USRP N210s with SBX daughterboards. Each end node is one USRP connected to a PC through an Ethernet cable. The NCMA AP has two USRPs connected through one MIMO cable so that the AP behaves like one node with two antennas. For the uplink channel, the AP sends beacon frames to trigger the three end users' simultaneous transmissions. We performed controlled experiments to simulate collisions of three users in random access networks. Our experiments were carried out at 2.585GHz center frequency with 5MHz bandwidth.

To benchmark our symbol-splitting rate-diverse NCMA system, we consider the following three systems:
\begin{enumerate}\leftmargin=0in
\item \emph{Rate-identical NCMA system} \\
This is the system based on the previous MIMO-NCMA system \referred{MIMONCMA_Globecom}\cite{MIMONCMA_Globecom} and it serves as a benchmark here. We extend the system in \referred{MIMONCMA_Globecom}\cite{MIMONCMA_Globecom} to support three users. The three-user PNC and MUD decoders discussed in Section \ref{sec:overview3} are used here. PHY-layer and MAC-layer bridgings are performed to increase system throughputs. Both BPSK and QPSK modulations are considered. The three users either all use BPSK or all use QPSK.

\item \emph{Direct Rate-Diverse NCMA system (DR-NCMA)} \\
This is the rate-diverse NCMA system directly generalized from the rate-identical system (DR-NCMA). We study the case where two users A and B adopt BPSK, and one user C adopts QPSK. As discussed in Section \ref{sec:ratediverse_ncma1}, DR-NCMA has only one PNC decoder (to decode $C^A \oplus C^B$). In particular, only users A and B are involved in PHY-layer and MAC-layer bridgings, and the performance of user C depends on the MUD decoder.

\item \emph{Symbol-splitting Rate-Diverse NCMA system (SR-NCMA)} \\
This is the rate-diverse NCMA system with symbol-splitting encoding for high-order modulated packets (SR-NCMA). Users A and B use BPSK, and user C uses QPSK. We implemented the symbol-splitting encoding scheme and the corresponding SR-NCMA PHY-layer decoders. All users can exploit PHY-layer and MAC-layer bridgings to improve throughputs.
\end{enumerate}

\subsection{Experimental Results} \label{sec:Exp2}
We evaluate the total system throughput and individual end users' throughputs: we first compare the PHY-layer and MAC-layer performances of SR-NCMA and DR-NCMA. After that, we evaluate the throughputs of rate-identical NCMA and rate-diverse NCMA.

We performed controlled experiments for different received SNRs, and calculated SNRs using the method in \referred{HalperinSNR10}\cite{HalperinSNR10}. The received powers of signals from end users A and B at the AP were adjusted to be approximately balanced at 8dB (we remark that the powers of each user could be slightly different due to channel fading, and the SNR presented here is the average SNR of all the received packets). For user C, we varied the SNR values from 8 to 14dB. For each SNR, the AP sent 1,000 beacon frames to trigger simultaneous transmissions of the three end users.

For the calculation of throughputs, we normalize one QPSK packet to two BPSK packets. The normalized throughputs for user $s$ and the whole NCMA system, $T{h^s}$ and $T{h^{sys}}$, are defined as follows:
\begin{align}
&T{h^s} = \frac{{{L_s} \times {N_s}}}{{{N_{Beacon}}}},{\rm{   }}s \in \{ A,B,C\}, \\
&T{h^{sys}} = \sum\limits_{s \in \{ A,B,C\} } {T{h^s}} ,
\label{equ:throughput}
\end{align}
where ${N_s}$ is the number of messages of user $s$ that have been recovered. ${N_{Beacon}}$ is the number of beacons, and ${L_s}$ is the number\footnote{In NCMA, the MAC-layer RS code's parameter $L$ (see Section \ref{sec:overview2}) can be different for different users. We choose an asymmetric case where ${L_C} = 2{L_B} = 4{L_A} = 32$ to achieve a better performance at MAC-layer bridging, based on our prior experimental results, and detailed explanation and justification for using asymmetric $L_s$ can be found in \referred{NCMA1}\cite{NCMA1}.} of normalized BPSK packets the AP needs in order to decode message ${M^s}$.

\subsubsection{Throughput comparisons between SR-NCMA and DR-NCMA} \label{sec:Exp21}
\begin{figure}[t]
\centering
\includegraphics[width=0.35\textwidth]{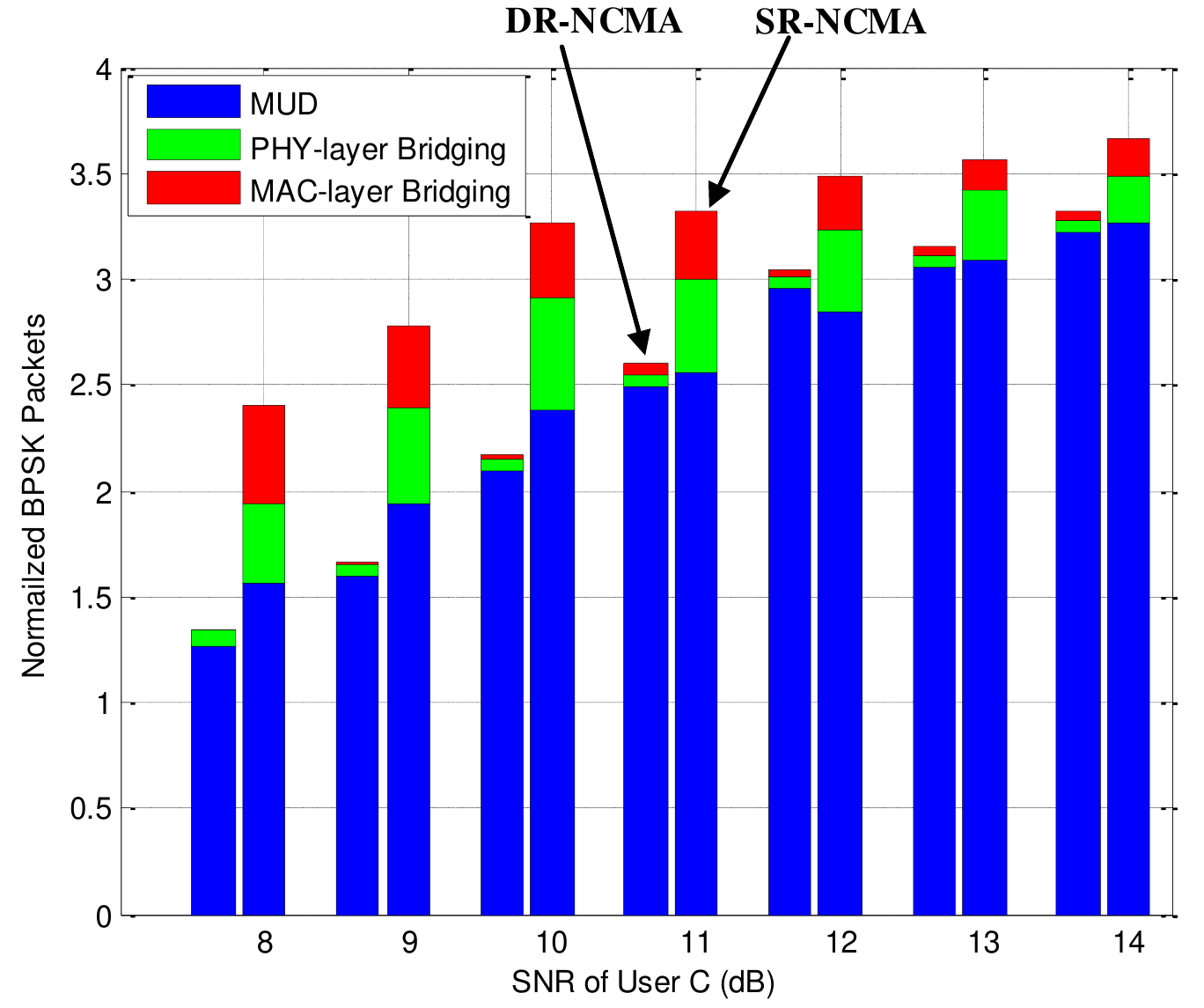}
\caption{Rate-diverse NCMA scheme throughput comparisons: SR-NCMA and DR-NCMA. The blue portions represent the throughputs when only MUD decoders are used; the green and red portions represent the extra throughputs when PNC decoders are also used with PHY-layer bridging and MAC-layer bridging, respectively. At the MAC layer, RS codes with ${L_C} = 2{L_B} = 4{L_A} = 32$ are used. The SNRs of A and B are fixed at 8dB while the SNR of C varies from 8dB to 14dB. }
\label{fig:rate_diverse_exp}
\end{figure}

We now compare the total system throughputs of two rate-diverse NCMA schemes, namely, SR-NCMA and DR-NCMA. We first evaluate the throughputs when only MUD decoders are used. Then, we study the system where PNC decodes are also used and where PHY-layer bridging is invoked.  After that, the overall throughputs with MAC-layer bridging are considered. The performance details are shown in Fig. \ref{fig:rate_diverse_exp}.

\textbf{(a)	MUD Decoders Performance:}
In Fig. \ref{fig:rate_diverse_exp}, the blue portions represent the throughputs of SR-NCMA and DR-NCMA when only MUD decoders are used. For MUD decoders, a key difference between the two schemes is the decoding of the QPSK packets of user C. In DR-NCMA, one MUD decoder tries to decode the whole QPSK packet $C^C$; while in SR-NCMA, two MUD decoders try to decode packets $C^{{C_I}}$ and $C^{{C_Q}}$. When user C's SNR is low (e.g., 8dB in Fig. \ref{fig:rate_diverse_exp}), it is likely that DR-NCMA fails to decode the whole QPSK packet $C^C$, but it is possible for SR-NCMA to decode one of the two packets $C^{{C_I}}$ or $C^{{C_Q}}$. Hence, we see that the MUD performance of SR-NCMA is better than DR-NCMA when user C has low SNRs. As user C's SNR increases, the MUD performances of these two schemes converge.

\textbf{(b)	PHY-layer Bridging Performance:}
A distinguishing feature of NCMA is the use of PNC packets to improve system throughputs by PHY-layer bridging and MAC-layer bridging. We now evaluate the extra throughput gain due to PHY-layer bridging (the green portions in In Fig. \ref{fig:rate_diverse_exp}). Since DR-NCMA has only one PNC decoder that decodes $C^A \oplus C^B$, and the QPSK user is not involved in PNC decoding, PHY-layer bridging yield little improvement in DR-NCMA (i.e., PHY-layer bridging can only happen between users A and B). However, for SR-NCMA, thanks to \emph{symbol-splitting encoding}, PHY-layer bridging can also happen between the QPSK user C and the BPSK users A and B. With PHY-layer bridging, SR-NCMA can have around 17\% throughput improvement over that with MUD decoders only.

\textbf{(c)	MAC-layer Bridging Performance:}
We now evaluate the throughput gain due to MAC-layer bridging (the red portions in Fig. \ref{fig:rate_diverse_exp}). Similar to the performance of PHY-layer bridging, MAC-layer bridging improves the performance of DR-NCMA very little because of the lack of PNC packets. In contrast, MAC-layer bridging can further improve the throughput of SR-NCMA by around 12\%. Therefore, the total system throughput of SR-NCMA is 40\% over that of DR-NCMA on average, as shown in Fig. \ref{fig:rate_diverse_exp}.

\begin{figure}[t]
\centering
\includegraphics[width=0.49\textwidth]{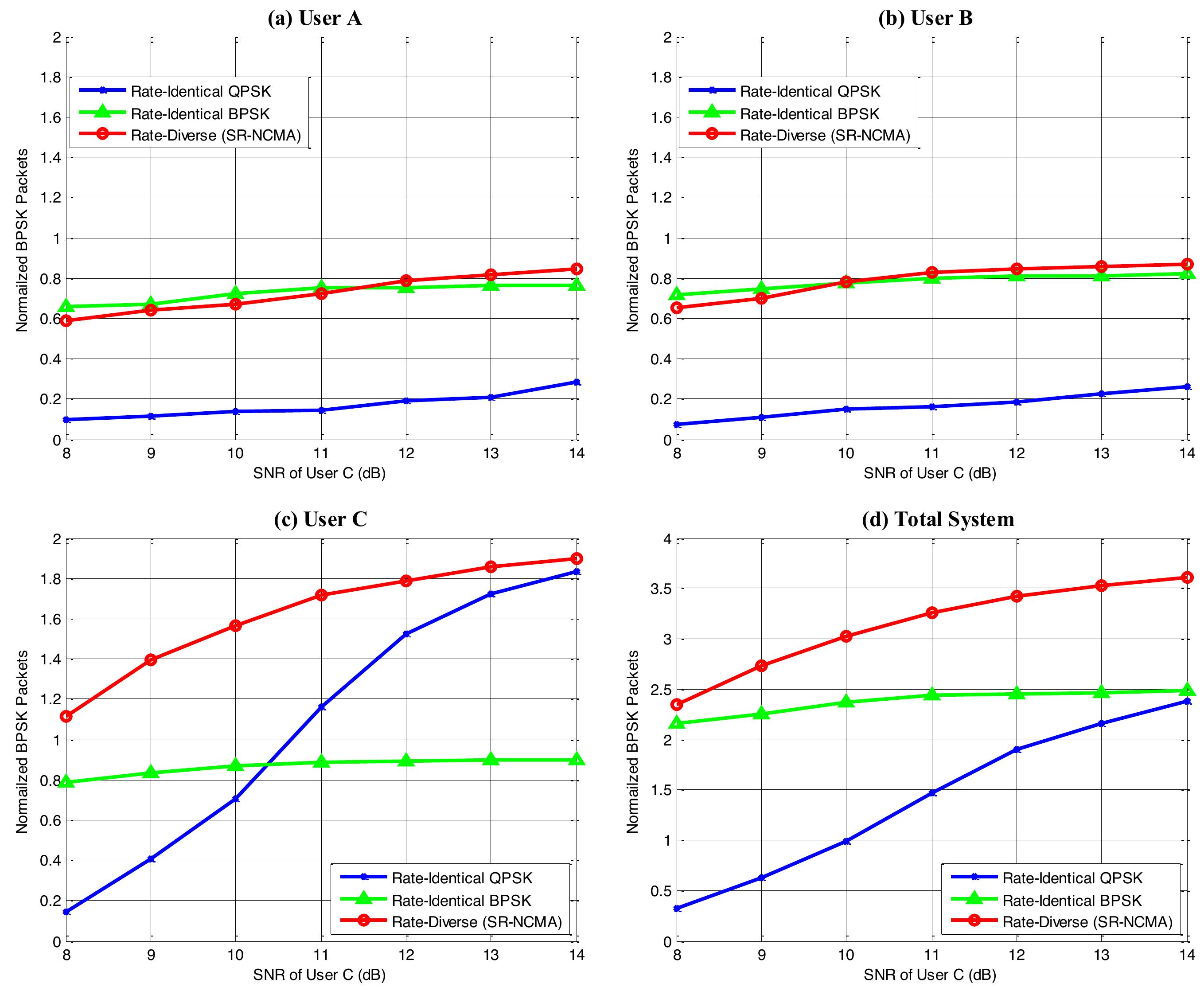}
\caption{Individual users' and total system's normalized throughput comparisons between rate-identical NCMA and rate-diverse NCMA. At the MAC layer, RS codes with ${L_C} = 2{L_B} = 4{L_A} = 32$  are used. The SNRs of A and B are fixed at 8dB while the SNR of C varies from 8dB to 14dB. }
\label{fig:exp_throughput}
\end{figure}

\subsubsection{Throughput comparisons between Rate-diverse NCMA and Rate-identical NCMA} \label{sec:Exp22}
We now compare the throughputs of rate-diverse NCMA and rate-identical NCMA. Since SR-NCMA performs better than DR-NCMA, we use SR-NCMA as the representative of rate-diverse NCMA in the rest of this section.

Let us focus on the rate-identical NCMA systems first. For individual users, we can see from Fig. \ref{fig:exp_throughput}(a)(b) that when QPSK is used, users A and B have low throughputs because of their low SNRs (fixed at 8dB while the SNR of user C varies from 8dB to 14dB). But when BPSK is used, although the throughputs of A and B improve, the throughput of user C is upper bounded by one BPSK packets per time slot even as SNR increases (i.e., user C is forced to use a low modulation order, not leveraging its high SNR to obtain better throughput), as shown in Fig. \ref{fig:exp_throughput}(c). Overall, the total system throughputs of both QPSK and BPSK rate-identical NCMA systems are bounded to no more than 3 normalized BPSK packets (see Fig. \ref{fig:exp_throughput}(d)). 

In contrast, the rate-diverse NCMA system, namely SR-NCMA, can allow users to choose their modulation order based on different channel conditions. From Fig. \ref{fig:exp_throughput}, we can see that the throughputs of BPSK users A and B is comparable to those in BPSK rate-identical NCMA, and at the same time, user C can achieve one QPSK packet per time slot at high SNRs. Overall, SR-NCMA achieves the highest total system throughput as shown in Fig. \ref{fig:exp_throughput}(d). For example, when user C's SNR is 12dB, the throughput of SR-NCMA is higher than those rate-identical NCMA systems operated with BPSK and QPSK by 40\% and 80\%, respectively.

\section{Conclusions}\label{sec:Conclusions}
We have developed a three-user rate-diverse NCMA system, where different users may use signal modulations commensurate with their respective channel SNRs. A key technology put forth by us to enable rate-diverse channel-coded PNC is \emph{symbol-splitting encoding}. Experiments on our software-defined radio prototype indicate that, compared with rate-homogeneous NCMA, rate-diverse NCMA can better exploit the varying channel conditions among users to achieve higher individual throughputs and higher overall system throughput in real wireless environment. Specifically, the system throughput of rate-diverse NCMA with BPSK+QPSK modulations outperforms those of rate-homogeneous NCMA where all users adopt BPSK and all users adopt QPSK  by 40\% and 80\%, respectively.

\ifCLASSOPTIONcompsoc
  \section*{Acknowledgments}
\else
  \section*{Acknowledgment}
\fi

The work of H. Pan and S. C. Liew was supported by the General Research Funds (Project No. 14204714) established under the University Grant Committee of the Hong Kong Special Administrative Region, China.
The work of L. Lu was partially supported by AoE grant E-02/08, established under the University Grant Committee of the Hong Kong Special Administrative Region, China, and partially supported by the NSFC (Project No. 61501390).


\bibliographystyle{IEEEtran}
\bibliography{mu_ncma}

\end{document}